\documentclass{aastex631}

\shorttitle{On the recent discovery claim of a new $z>7$ quasar}
\shortauthors{Bosman, Davies \& Ba\~{n}ados}

\graphicspath{{../plots}{figures/}{./}}

\begin{document}

\title{On the discovery claim of a new $z>7$ quasar}
%On the nature of the recently-discovered "quasar" at z>7
%Is J0816+2134 actually a quasar at z>7?

\author[0000-0001-8582-7012]{Sarah E. I. Bosman}
\affiliation{Max-Planck-Institut für Astronomie\\
Königstuhl 17, D-69117 \\
Heidelberg, Germany}

\author[0000-0003-0821-3644]{Frederick B. Davies}
\affiliation{Max-Planck-Institut für Astronomie\\
Königstuhl 17, D-69117 \\
Heidelberg, Germany}

\author[0000-0002-2931-7824]{Eduardo Ba\~{n}ados}
\affiliation{Max-Planck-Institut für Astronomie\\
Königstuhl 17, D-69117 \\
Heidelberg, Germany}

\begin{abstract}
\citet{Koptelova} (K22) recently claimed a new quasar discovery at $z=7.46$. After careful consideration of the publicly-available data underlying K22's claim, we find that the observations were contaminated by a moving Solar System object, 
% most
likely a main-belt asteroid. In the absence of the contaminated photometry, there is no evidence for the nearby, persistent WISE source being a high-redshift object; in fact, a detection of the source in DELS $z$-band rules out a redshift $z>7.3$. We present our findings as a cautionary tale 
%of the potential for passing asteroids to contaminate photometric selections.
of the dangers of passing asteroids for photometric selections.
\end{abstract}


\section{Introduction}

Luminous quasars at the highest redshifts have far-ranging uses in astrophysics and cosmology, from studying reionisation to the origin of supermassive black holes \citep{Fan-review,Bosman22,Eilers21}. 
%They provide the currently strongest constraints on the end of hydrogen reionisation \citep{Bosman22,Zhu21}, early metal enrichment of the CGM and the IGM \citep{Cooper19,Suksien22}, the origin and growth of supermassive black holes \citep{Eilers21, Lai22}, and the growth of the first massive dusty galaxies \citep{Venemans12,Walter22}, among many others. 
As such, the search for new quasars at the high-redshift frontier, currently at $z\gtrsim7$, is a very active and competitive field (e.g.~\citealt{Wang21}).

\section{Claim of discovery a new quasar at $z\simeq7.5$}

Recently, a pre-print was posted on the arXiv server claiming the discovery of a new quasar at $z=7.46$ (K22). 
The candidate was initially selected based on survey photometry from WISE \citep{WISE} and UKIDSS+UKIRT \citep{UKIDSS}. 
%Specifically, 
The identification was based on the following criteria: \begin{enumerate}
\item The candidate was detected in the WISE/W1 and WISE/W2 photometric bands, with W2 being brighter;
\item It was (thought to be) brighter in the UKIDSS/$J$ photometric band than in the W1 band, with a reported $J$-band magnitude of $J=19.3$ (all magnitudes AB);
\item It was undetected in all Pan-STARRS \citep{Pan-STARRS} bands $grizy$. 
%the UKIDSS/$Z$ band with $Z>22.8$.%, and very faint in UKIDSS/$Y$ with $Y=21.6$.
% FBD: ALERT ^ The above is not actually true! They used non-detections in all PS1 bands
% as their selection criterion, not UKIDSS/Z. 
% They double-checked UKIDSS Z and Y as a cross-check but it is not their original selection.
\end{enumerate}
Indeed, such a large difference between a bright $J$-band and a non-detection in the $z$-band, of over $3$ magnitudes, would justify the selection of the object as a potential $z>7$ quasar.

However, after analysing the public UKIRT/UKIDSS data of the candidate, we have uncovered evidence that the $J$ band photometry used by K22 is contaminated by a passing asteroid. 
%As can be seen in the top row of Figure 1, 
The top row of Figure 1 shows that 
% FBD: Below my version of the sentence, it is roughly the same length tho
the object centered in the $J$-band image is offset by about 2'' in the $Y$-band image observed $23$ minutes earlier.
%the candidate is offset by about 2'' between the time of the $J$-band observation and the $Y$-band roughly $23$ minutes earlier. 
By extrapolating the object's trajectory, we further located it
%as it was captured 
in the UKIDSS $H$ and $K$ band imaging 
%of UKIRT 
taken on the previous day about $1$'$22$'' away (rightmost two panels of Fig.~1). We obtained flux-weighted centroids for the $4$ known positions of the moving object with \textit{SourceExtractor} \citep{Bertin96}. We input those position-time coordinates into the web-based platform \textit{Find\_Orb}\footnote{\url{https://www.projectpluto.com/find_orb.htm}}, which performs orbit reconstruction for Solar System objects. The object's motion is entirely consistent with a heliocentric orbit with a semi-major axis of $2.5 \pm 1$ au, identifying it as a likely main-belt asteroid. Large uncertainties on orbital elements originate from the short time baseline between the $4$ chance observations and prohibit us from 
%further 
accurately extrapolating the motion beyond a few weeks.
%months.
%As far as we are aware, no other imaging survey was ongoing around November or December 2008 which would have serendipitously captured the object.
% FBD: My version of the above sentence:
We are not aware of any other imaging survey which would have serendipitously captured the object around November or December 2008.

\begin{figure*}
	\centering
    \includegraphics[trim=30 0 30 30,clip,width=1\textwidth]{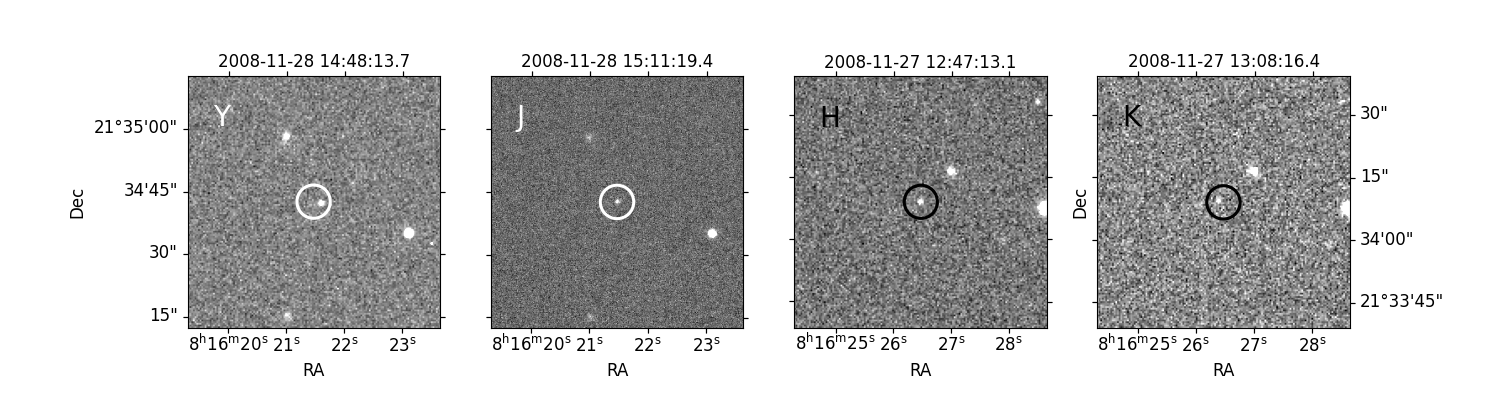}
    \includegraphics[trim=30 0 30 30,clip,width=1\textwidth]{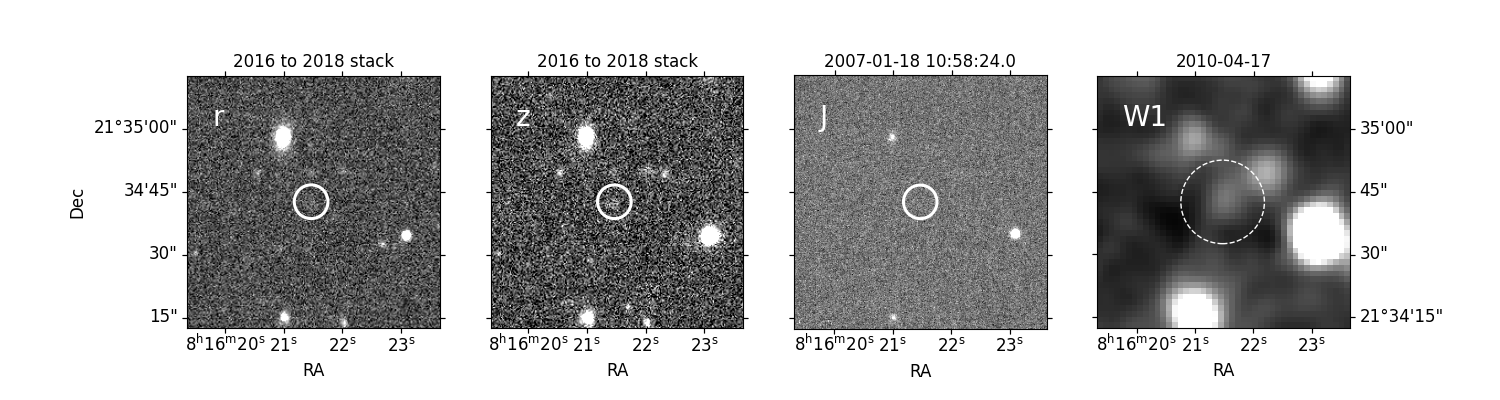}
    \caption{\textit{Top:} Motion of the contaminating asteroid in UKIDSS. 
    %Times are given 
    The observing epoch is indicated
    above each panel; the central circle is $8$'' across. The $H$ and $K$ band images are spatially offset from the $J$ and $Y$ bands by about $1$'$22$'' (W1 circle is $20$'' across). %The object's motion is consistent with a main-belt asteroid. 
\textit{Bottom}: photometry of the candidate object at times uncontaminated by the asteroid. The object is undetected in the UKIDSS $J$ band; %, and does not show the strong brightness drop towards the $z$ band characteristic of a high-redshift quasar; 
a detection in DELS $z$ band conclusively rules out a redshift $z>7.3$.}
    \label{fig:example}
\end{figure*}

Excluding the contaminated photometric frames, the object is undetected in UKIDSS $J$. % or $H$ (bottom rightmost panels of Fig.~1). 
However, a WISE source, detected in W1 and W2, is indeed  %very near
consistent with the location given by K22 (bottom panels of Fig.~1). 
%(2''). %; however, the coordinates of the candidate as given in K22 correspond to the centroid of the passing asteroid in the $J$ band, not to the persistant WISE source. I
%However, in the absence of the contaminated $J$-band photometry, 
%However, we find nothing to indicate that this WISE source could be a high-redshift quasar based on the UKIDSS photometry. 
%Finally, w
We examined this WISE source in the DESI Legacy Imaging Surveys (DELS; \citealt{Dey19}), which provide significantly deeper $z$-band imaging than both UKIDSS and Pan-STARRS. We find that the WISE source is detected at $>5\sigma$ in the $z$ band %, with a centroid closer to the location of the WISE photometry, as opposed to the asteroid contaminating the $J$ UKIDSS frame %and appears potentially extended 
(Fig.~1). A detection in the $z$-band completely rules out the object being a quasar at $z>7.3$, since full absorption by the neutral intergalactic medium would be unavoidable at rest-frame wavelengths $\lambda<1215.67$\AA.

As a follow-up to the photometric selection of the candidate, K22 present a spectrum of the object taken with the GNIRS spectrograph %on the Gemini telescope 
\citep{GNIRS2,GNIRS1}. The timing of the spectroscopic observations mean that the spectrum originates in the persistent WISE source, and are uncontaminated by the asteroid. This spectrum as shown in K22 
%supposedly 
% FBD: 
allegedly 
shows features confirming the object's nature as a $z=7.46$ quasar: a break in the continuum at $\lambda\sim1.1\mu$m, and a Mg~{\small{II}} emission line at $\lambda\sim2.35\mu$m. We reduced the publicly-available raw spectral data with the \textit{PypeIt} spectroscopic reduction package \citep{Pypeit-official} and can confirm neither of these features. While the WISE object is detected as a faint continuum, light is clearly present at $\lambda<0.95\mu$m: there is no sharp spectral break. 
This is in agreement with the detection of the object in DELS $z$-band. We do not see a clear emission-line feature at $\lambda\sim2.35\mu$m, nor any other wavelength.

%\section{Conclusion}

%

\bibliography{bibliography}
\bibliographystyle{aasjournal}

\end{document}